# Investigation of thermal stability of hydrogenated amorphous Si/Ge multilayers

A. Csik[1*], M. Serényi[2], Z. Erdélyi[3], A. Nemcsics[2], C. Cserhati[3], G.A. Langer[3], D.L. Beke[3], C. Frigeri[4], A. Simon[1]

[1]*Institute of Nuclear Research, Hungarian Academy of Sciences (ATOMKI), P.O. Box 51, Debrecen H-4001, Hungary*

[2]*MTA-MFA Institute, Konkoly-Thege ut 29-33, Budapest H-1121, Hungary*

[3]*Department of Solid State Physics, University of Debrecen, P.O. Box 2, Debrecen H-4010, Hungary*

[4]*CNR-IMEM Institute, Parco Area delle Scienze, 37/A, 43010 Parma, Italy*

**Abstract**

Thermal stability of hydrogenated amorphous Si/Ge multilayers has been investigated by Scanning Electron Microscopy (SEM), Transmission Electron Microscopy (TEM) and Small-Angle X-Ray Diffraction (SAXRD) techniques. Amorphous H-Si/Ge multilayers were prepared by RF sputtering with 1.5 and 6 ml/min $H_2$ flow-rate. It is shown by Elastic Recoil Detection Analysis (ERDA) that the hydrogen concentration increased by increasing $H_2$ flow-rate. Annealing of the samples was carried out at 400 and 450 °C for several hours. It has been observed that samples prepared with 6 ml/min flow-rate at both annealing temperatures underwent significant structural changes: the surface of the samples was visibly roughened, gas bubbles were formed and craters were created. The decay of the periodic structure of Si and Ge layers in these types of multilayers was faster than in non-hydrogenated samples. Samples prepared with 1.5 ml/min flow-rate have similar behaviour at 450 °C, but at 400 °C the decay of the first order SAXRD peaks was slower than in case of the non-hydrogenated multilayers. Qualitatively the observed behaviour can be explained by the fast desorption of the saturated hydrogen, leading to the formation of bubbles and craters at 450 $^{o}$C, as well as, at 400$^{o}$C in the sample with lower H-content, by the possible passivation of the dangling bonds resulting in a slowing down of the diffusion intermixing.

---

[*] Corresponding author. *Fax*: +36-52-416181
*E-mail address*: csik@atomki.hu (A. Csik)

**1. Introduction**

Hydrogenated amorphous silicon germanium alloys (a-Si$_{1-x}$Ge$_x$:H) are used in multi-junction solar cells in order to increase the efficiency of the cells [1,2]. One of the advantages of this material is the possibility to vary its band gap with the germanium concentration for the optimization the exploitation of the solar spectrum. In order to improve the stability of the solar cells, a multijunction structure is useful because each photovoltaic layer can be thinner than that of the single-junction solar cells. In case of Si/Ge amorphous multilayers the improvement of electronic properties is also accomplished by enriching them with hydrogen to passivate the dangling bonds of Si and Ge [3].

It is known that multilayers, as artificial compositionally modulated materials, are not equilibrium structures. In particular, they have high interfacial density gradients and sufficient atomic mobility even at moderate temperatures, hence changes in the composition profiles are expected to occur. Thus, not only analysis of the light induced structural changes (i.e. Staebler-Wronski effect) [4,5], but also investigation of the thermal stability and understanding of the factors controlling the structural changes of these multilayers is required. Exploration of these mechanisms is very important for the interpretation of operation and prediction of lifetime before integrating these structures into micro and opto-electronics devices. In our previous works the thermal stability of non-hydrogenated Si/Ge multilayers [6,7] was extensively studied, but there is no information available in literature on the effect of hydrogen addition in the intermixing of Si and Ge layers.

In this work we summarize our previous results on the structural changes [8,9], caused by the fast release of hydrogen, and present preliminary results on diffusion intermixing in amorphous Ge/Si multilayers prepared with different hydrogen content.

## 2. Experimental details

Hydrogenated and not-hydrogenated amorphous Si/Ge multilayers with 3 nm nominal layer thickness were deposited on (100) oriented Si substrates by the conventional radio frequency sputtering apparatus (Leybold Z 400) with basic vacuum pressure of $1 \cdot 10^{-6}$ mbar. During sputtering the chamber pressure was kept at $2 \cdot 10^{-2}$ mbar and the substrate was water cooled. A 1.5 kV wall potential was applied to the Si and Ge targets while argon (purity 99.999 %) was used as sputtering gas. The multilayers contained a sequence of 50 Si/Ge bi-layers. The hydrogenated Si/Ge samples have been deposited by adding hydrogen in the sputtering chamber at a flow-rate of either 1.5 or 6 ml/min. Both not-hydrogenated and hydrogenated multilayers have been annealed in high purity argon (99.999%) atmosphere at 400 and 450 °C.

After preparation the hydrogen content of the multilayer films was measured with Elastic Recoil Detection Analysis (ERDA) and it was clearly observed that (a) hydrogen absorption occurred and (b) the hydrogen content was larger for higher flow-rate. In order to check the possible hydrogen loss of the sample during the measurements, Rutherford backscattering spectrometry (RBS) was also applied simultaneously. The ERDA data were normalized to the RBS data. Micro-ERDA and micro-RBS with helium beam were performed on the Oxford-type nuclear microprobe facility at ATOMKI, Debrecen [10]. The energy of the $^4$He$^+$ ion beam was 1600 keV. Three ORTEC-type surface barrier silicon detectors (50 mm$^2$ sensitive area and 18 keV system energy resolution) were used for micro-ERDA and micro-RBS experiments simultaneously. The ERDA detector was placed at a recoil angle of 30° (IBM geometry) mounted with a 6 μm thick Mylar absorber and 1.1 mm wide vertical aperture. The tilt angle of the sample was 75°. Two detectors collected the micro-RBS data; one of them was placed at a scattering angle of 165° at Cornell geometry and the other one was set to 135° at IBM geometry. The total deposited charge was collected on the sample.

The investigations of the samples were carried out by Scanning Electron Microscopy – SEM (type Hitachi S-4300 CFE) operated in the secondary electron mode, Transmission Electron Microscopy – TEM (type JEOL 2000) and Small-Angle X-Ray Diffraction instrument equipped with Siemens Cu-anode X-Ray tube. The TEM specimens were prepared by mechanical thinning of sandwiches, containing a small piece of the sample, down to 30 μm followed by Ar ion beam thinning down to electron transparency. SEM was used for surface morphological examinations.

One of the powerful and widely used techniques for characterization of multilayers is the Small Angle X-ray Diffraction (SAXRD) [11]. It is a non-destructive method and provides structural information on the atomic scale. The Θ−2Θ SAXRD scans were employed to investigate the diffusion processes by measuring the intensity decrease of the first order Bragg-peak which can arise from the chemical modulation of the structure [12]. Since the intensities of the low angle Bragg-peaks are strongly influenced by the sharpness of the interfaces, the decay of the intensity of the first order peak in small angle X-ray diffraction gives information about the intermixing of the Si and Ge layers. This behaviour can even be used to measure the interdiffusion coefficients in multilayers. In fact the slope of the logarithm of the normalized intensity of the first order peak versus the annealing time is inversely proportional to the diffusion coefficient.

Two types of hydrogenated amorphous Si/Ge multilayers, prepared with 1.5 and 6 ml/min flow-rates and annealed at 400 and 450 $^{o}$C for different annealing times, were investigated.

## 3. Results and discussion

After diffusion heat treatments – as it was discussed in our previous papers [8,9] – it was observed that the hydrogenated Si/Ge multilayers underwent significant structural changes. The originally flat surfaces of the samples, prepared with 6 ml/min flow-rate, were visibly roughened at both applied annealing temperatures; gas bubbles were formed and craters were created (Fig.1a.). However the samples prepared with lower $H_2$ concentration (with 1.5 ml/min flow-rate of $H_2$) showed a similar effect only at 450 $^{o}$C. During annealing at 400 $^{o}$C the surface of these samples remained smooth (no craters) and only moderate bubble formation has occurred (Fig.1b.). This indicates that the structure of the samples prepared with low hydrogen concentration were more stable at 400 $^{o}$C, but at higher annealing temperature (450 $^{o}$C) they can degrade similarly as samples with higher hydrogen content (prepared with 6 ml/min $H_2$ flow-rate).

The gas bubbles at 450 $^{o}$C are formed due to the enhanced precipitation of the supersaturated hydrogen. The intensive growth of the bubbles can eventually locally blow up the multilayered film. On the other hand, the TEM pictures (Fig.2.) and small angle X-ray measurements show that despite the formation of craters, the multilayer structure is preserved. This offers an opportunity to perform a diffusion experiment – able to indicate classical intermixing between amorphous Ge and Si – still containing some hydrogen. Thus, after each annealing the small angle X-ray spectrum has been recorded and the logarithm of the normalized intensity of the first order Bragg-peak has been plotted as the function of an annealing time. Figure 3. shows these experimental results. The slope of the curves can show that diffusion took place between the Si and Ge layers. The non-linearity of the time dependence of $lnI/I_o$ indicates that the diffusion coefficient strongly depends on the composition and/or significant stresses of diffusion origin were formed during the heat treatments [7]. Nevertheless some qualitative conclusions can be obtained from the comparison of the above decay curves. First, both hydrogenated samples show a fast structural degradation at 450 $^o$C and, according to the SEM pictures, beside the diffusion intermixing the morphological changes (bubbles, craters) can also play a significant role. The same conclusion can be drawn from the decay curve at 400$^o$C with high hydrogen content. On the other hand, interestingly the slope (behind the first fast drops) of the curve for the sample produced with low $H_2$ flow-rate and annealed at 400 $^o$C is slower than the corresponding slope of the not-hydrogenated sample. It was also observed that after initial fast formation of bubbles there were no more structural changes and probably only atomic scale intermixing took place. One can suppose that in this case the different slopes of the decay curves indicate the difference in the rate of the diffusional intermixing in hydrogenated and not-hydrogenated samples.

It is known that hydrogen inactivates the dangling bonds which are well known to introduce deep gap states thus improving the electrical and optical properties of the materials [13,14]. Although, the diffusion mechanism in amorphous semiconductors is not clearly understood yet, it is believed that dangling bonds play an important role [15]. Thus it is expected that the presence of the hydrogen modifies the diffusion processes apparently.

There is experimental evidence that the diffusion is faster in amorphous Ge than in amorphous Si [16] and bond energy of Ge-H is also lower than the Si-H one [5,17]. Thus it can be assumed that during annealing the hydrogen release is faster in the Ge layers. At high temperature (450 $^o$C) this yields formation of small bubbles here and their subsequent coalescence destroys a significant fraction of the periodic modulation in the multilayer structure. This is obviously manifested in fast decay of the first order SAXRD peaks.

On the other hand the decay curve obtained at 400 $^o$C on the sample with lower hydrogen content indicates that the above macroscopic degradation is more moderate and restricted to the very first stage of the process. Consequently the experimental fact that the diffusion intermixing is slower in this sample as compared to the not-hydrogenated one (the slopes of the decay curves at longer annealing times are different accordingly) can be interpreted as due to the still remaining hydrogen. This can inactivate the dangling bonds and decrease the diffusion coefficient.

Obviously to give a more comprehensive and unequivocal interpretation, it is necessary to get more extended and accurate experimental information about the hydrogen concentration of the samples before and after heat treatments. Thus further experiments are in progress in order to determine the exact initial hydrogen concentration, to clear it up how the diffusion process depends indeed on the hydrogen content left after the fast initial stage.

## 4. Conclusion

Hydrogenated and not-hydrogenated amorphous Si/Ge multilayers, deposited by RF magnetron sputtering under different hydrogen flow-rates, have been investigated by SEM, TEM, SAXD and ERDA methods. It has been observed that samples prepared under 6 ml/min hydrogen flow-rate at both annealing temperatures (400 and 450 $^o$C) underwent significant structural changes. The overall degradation of the periodic structure in these types of the multilayers was faster than in not-hydrogenated samples. On the other hand, diffusion intermixing at 400 $^o$C, in samples prepared in lower hydrogen flow-rate (1.5 ml/min), was slower than in the not-hydrogenated ones. In this case the macroscopic degradation by formation of bubbles and craters was more moderate as compared to the results obtained at 450$^o$C or in samples with higher hydrogen content and was restricted to the very first stage of the decay. We suggest that the still remaining hydrogen can inactivate the dangling bonds of Si and Ge and the diffusion intermixing slows down until hydrogen remains in the layers.

**Acknowledgements**

This work was supported by the Scientific Cooperation Agreement between HAS (Hungary) and CNR (Italy) as well as by the EU co-funded Economic Competitiveness Operative Programme (GVOP-3.2.1.- 2004-04-402/3.0) and by Hungarian Research Fund OTKA grant No. K-61253, NK73424.

**Figure captions**

Figure 1. Secondary electron image of the hydrogenated sample prepared with 6 ml/min hydrogen flow-rate after 8 hours annealing at 450 $^o$C (a.) and sample prepared under 1.5 ml/min hydrogen flow-rate after 10 hours annealing at 400 $^o$C (b.)

Figure 2. TEM image of an annealed hydrogenated sample (hydrogen flow-rate 6ml/min, annealing temperature 450 $^o$C)

Figure 3. Decay of the intensity of the first order peak as a function of annealing time for the samples annealed at 400 $^o$C (a.) and 450 $^o$C (b.)

Figure 1. A. Csik, M. Serényi, Z. Erdélyi, A. Nemcsics, C. Cserhati, G.A. Langer, D.L. Beke, C. Frigeri, A. Simon

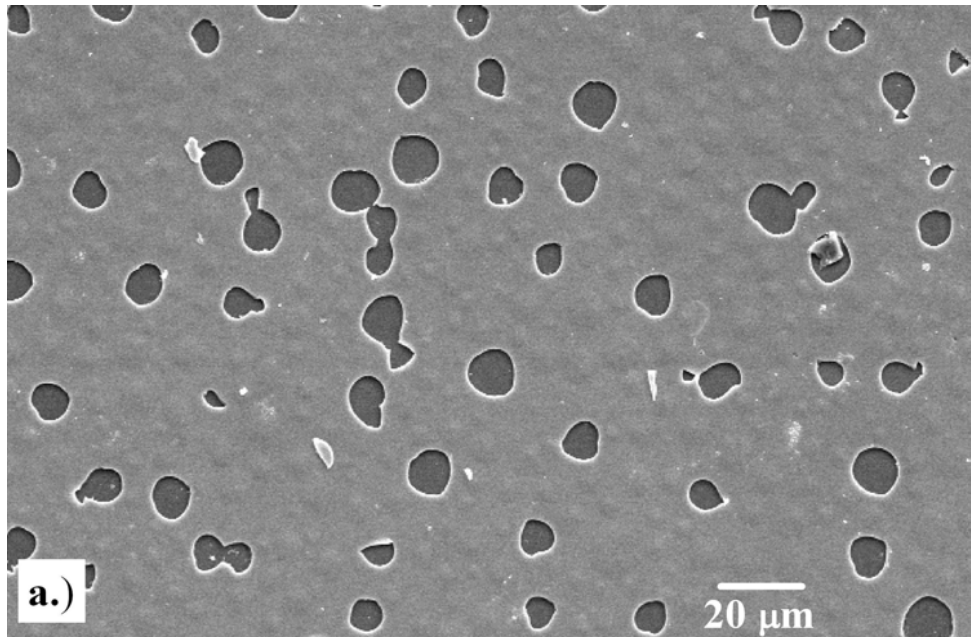

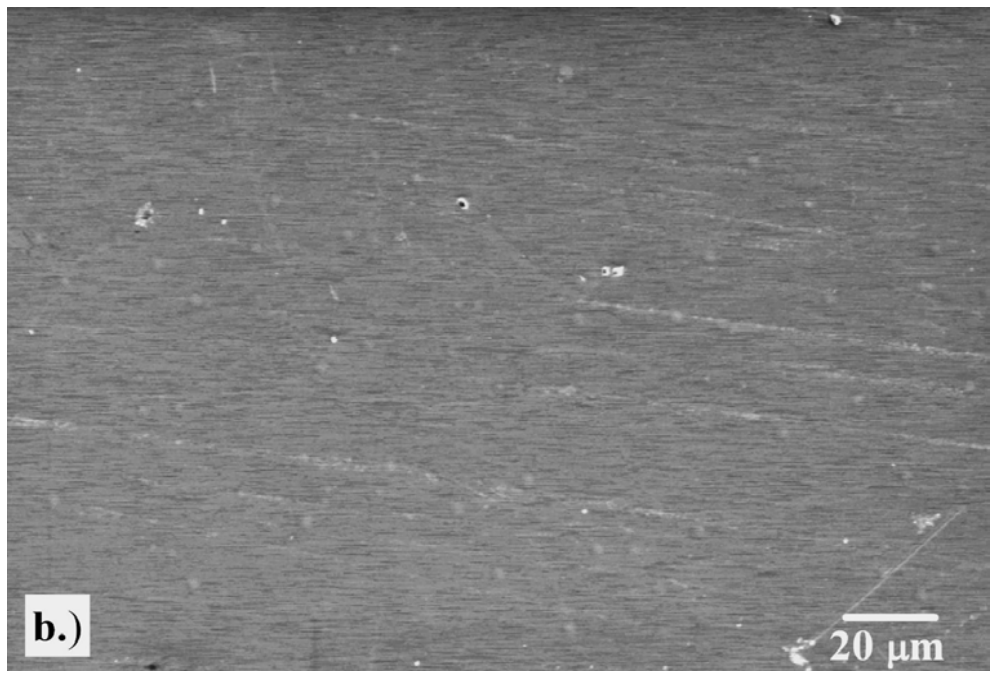

Figure 2. A. Csik, M. Serényi, Z. Erdélyi, A. Nemcsics, C. Cserhati, G.A. Langer, D.L. Beke, C. Frigeri, A. Simon

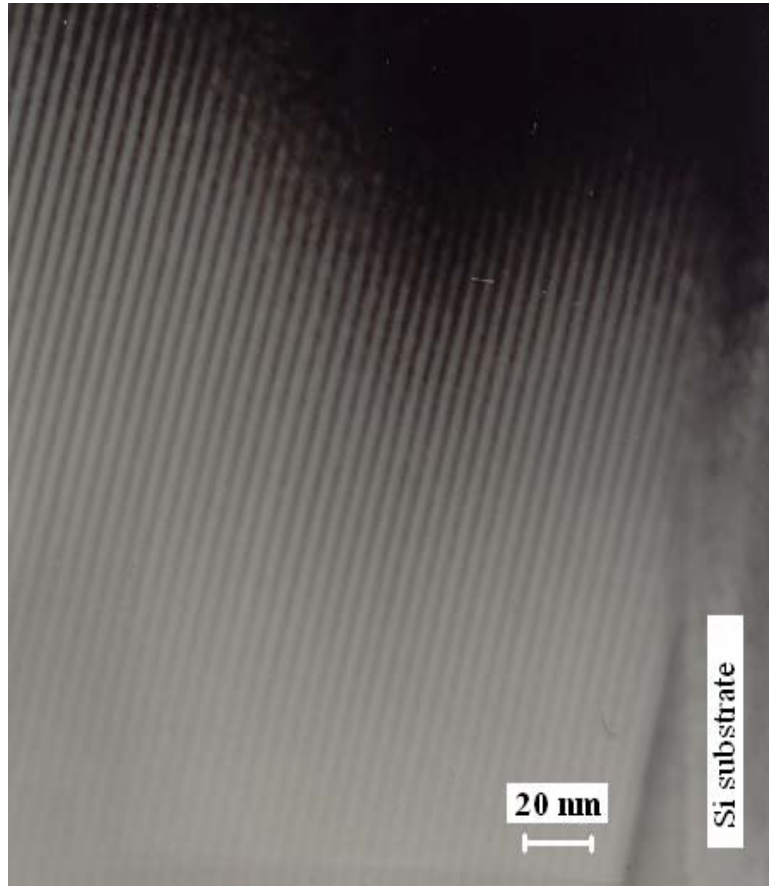

Figure 3. A. Csik, M. Serényi, Z. Erdélyi, A. Nemcsics, C. Cserhati, G.A. Langer, D.L. Beke, C. Frigeri, A. Simon

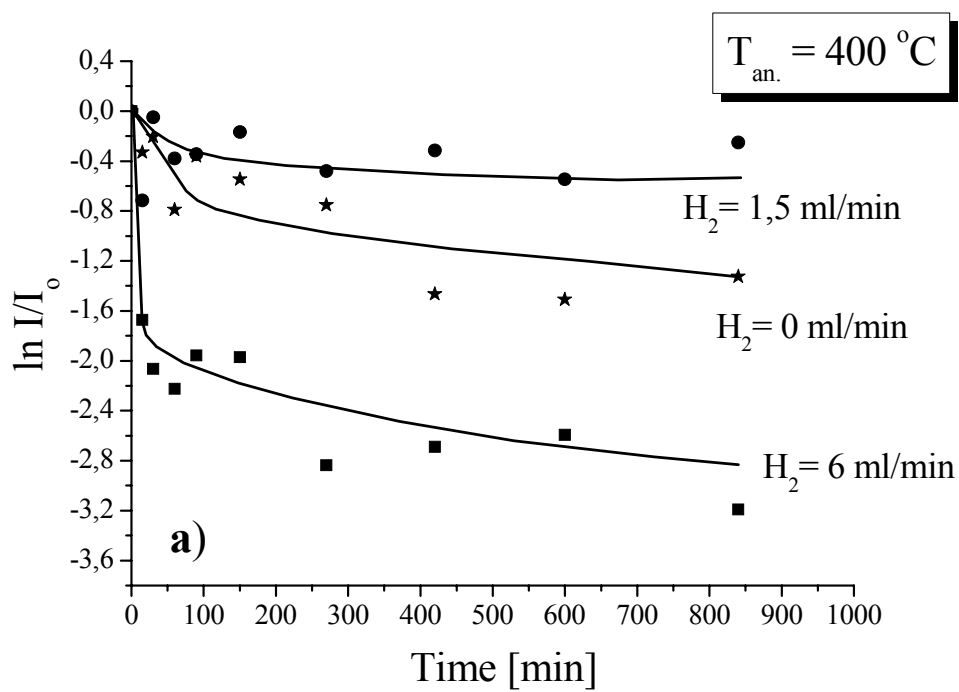

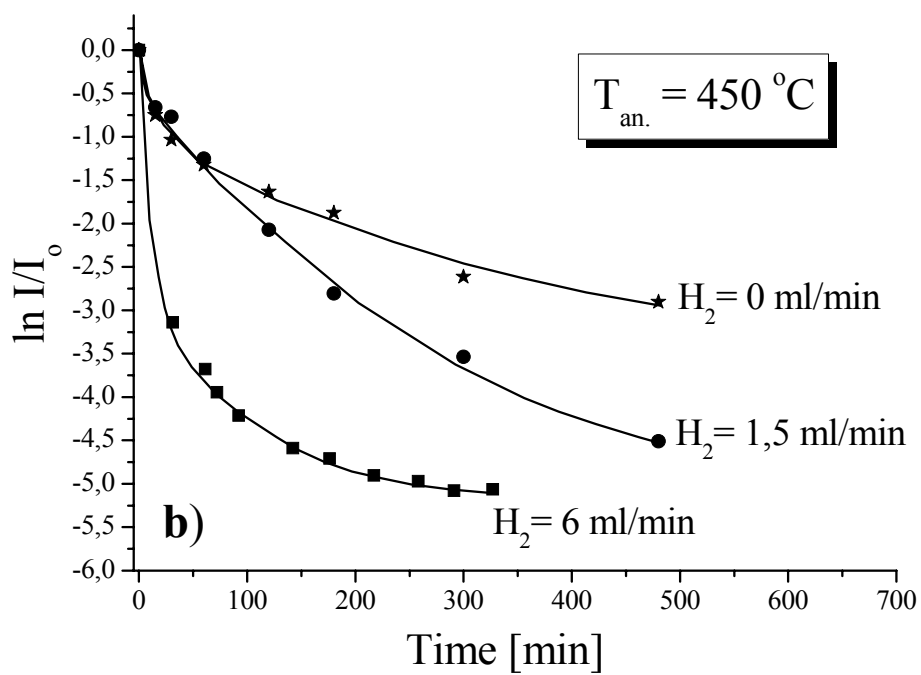